\documentclass[useAMS,usenatbib]{mn2e}

\usepackage{latexsym,graphicx,natbib}
\usepackage[english]{babel}
\usepackage{mciteplus}

\usepackage{color}

\newcommand\kms{{\rm\,km\,s^{-1}}}
\newcommand\msun{\rm\,M_\odot}

\def\apgt{\ {\raise-.5ex\hbox{$\buildrel>\over\sim$}}\ }
\def\aplt{\ {\raise-.5ex\hbox{$\buildrel<\over\sim$}}\ }

\title[WR\,120bb and WR\,120bc: a pair of WN9h stars with possibly interacting circumstellar shells]{WR\,120bb and WR\,120bc: a pair of WN9h stars with possibly interacting circumstellar shells}

\author[S. Burgemeister et al.]
{S.~Burgemeister$^{1}$\thanks{E-mail:
sburgem@astro.physik.uni-potsdam.de (SB);        
vgvaram@mx.iki.rssi.ru (VVG); Guy.Stringfellow@colorado.edu (GSS);
akniazev@saao.ac.za (AYK); htodt@astro.physik.uni-potsdam.de (HT); wrh@astro.physik.uni-potsdam.de (WRH)},
V.~V.~Gvaramadze$^{2,3}$\footnotemark[1],
        G.~S.~Stringfellow$^{4}$\footnotemark[1]\thanks{Observations obtained with the Apache Point Observatory 
3.5-meter telescope, which is owned and operated by the Astrophysical 
Research Consortium.},
	A.~Y.~Kniazev$^{5,6,2}$\footnotemark[1],
\newauthor	        H.~Todt$^1$\footnotemark[1],
        and W.-R.~Hamann$^{1}$\footnotemark[1]\\
$^{1}$Institute for Physics and Astronomy, University Potsdam, 14476
     Potsdam, Germany\\
$^{2}$Sternberg Astronomical Institute, Lomonosov Moscow State
      University, Universitetskij Pr. 13, Moscow 119992, Russia\\
$^{3}$Isaac Newton Institute of Chile, Moscow Branch, Universitetskij
      Pr. 13, Moscow 119992, Russia\\
$^{4}$Center for Astrophysics and Space Astronomy, University of
      Colorado, 389 UCB Boulder, Colorado 80309-0389, USA\\
$^{5}$South African Astronomical Observatory, PO Box 9, 7935
      Observatory, Cape Town, South Africa\\
$^{6}$Southern African Large Telescope Foundation, PO Box 9, 7935
      Observatory, Cape Town, South Africa
              }

\begin{document}

\date{}

\pagerange{\pageref{firstpage}--\pageref{lastpage}} \pubyear{2012}

\maketitle

\label{firstpage}

\begin{abstract}

Two optically obscured Wolf-Rayet (WR) stars have been recently
discovered by means of their infrared (IR) circumstellar shells,
which show signatures of interaction with each other. Following
the systematics of the WR star catalogues, these stars obtain the
names WR\,120bb and WR\,120bc. In this paper, we present and
analyse new near-IR, $J$, $H$, and $K$-band, spectra using the
Potsdam Wolf-Rayet (PoWR) model atmosphere code. For that purpose,
the atomic data base of the code has been extended in order to
include all significant lines in the near-IR bands.

The spectra of both stars are classified as WN9h. As their spectra are
very similar the parameters that we obtained
by the spectral analyses hardly differ. Despite their late
spectral subtype, we found relatively high stellar temperatures of
63 kK. The wind composition is dominated by helium, while hydrogen
is depleted to 25 per cent by mass. 

 Because of their location 
in the Scutum-Centaurus arm, WR\,120bb and WR\,120bc appear highly 
reddened, $A_{K_{\rm s}} \approx 2$ mag.
We adopt a common distance of 5.8\,kpc to 
both stars, which complies with the typical absolute $K$-band 
magnitude for the WN9h subtype of $-6.5$ mag, is consistent with their observed extinction based on comparison with other massive stars in the region,  and allows for the possibility that their shells are interacting with each other. This leads to 
luminosities of $\log(L/L_\odot ) = 5.66$ and 5.54 for WR\,120bb 
and WR\,120bc,  with large uncertainties due to the adopted distance. 

The values of the luminosities of WR\,120bb and WR\,120bc imply that 
the immediate precursors of both stars were red supergiants (RSG).
This implies in turn that the circumstellar shells associated with
WR\,120bb and WR\,120bc were formed by interaction between the
WR wind and the dense material shed during the preceding RSG phase.

\end{abstract}


\section{Introduction}

Wolf-Rayet (WR) stars play a key role in the cosmic circuit of
matter. As they are massive hot stars with a strong stellar wind,
they continuously enrich their environment with metals and ionise
the surrounding interstellar matter. Furthermore, they are
considered as progenitors of supernovae and $\gamma$-ray bursts.
Nevertheless, the origin of WR stars is not yet safely established
and their physics is far from being fully understood. Detection
and study of new WR stars is therefore warranted.

Until recently, the main channel for detection of WR stars was
through narrow-band optical surveys \citep[e.g.][and
references therein]{1999AJ....118..390S}. The advent of infrared (IR) observations
brought new possibilities. Searches for new WR stars using
narrow-band mid-IR surveys \citep{2009AJ....138..402S,2012AJ....143..149S} and IR
colour-based selection of objects  \citep{hadfield2007,2011AJ....142...40M}
led to the almost two-fold increase of the
known Galactic population of these stars.

Another important tool for revealing WR and other types of evolved
massive stars is through detection of their circumstellar nebulae.
Classically it is known that some WR stars exhibit ring nebulae
that are emitting in H$\alpha$ and other optical lines 
\citep{1965ApJ...142.1033J,1983ApJS...53..937C,1994ApJS...93..455D,1995AJ....109.1839M}.
Because of the high obscuration in the Galactic plane 
(where most sites of massive star formation are located) the distant 
WR nebulae can be detected mainly through their mid- and far-IR 
emission
\citep[e.g.][]{1988ApJ...329L..93V,1991ApJ...366..181M,1992ApJ...384..197M},
 whose origin can be attributed to radiatively heated 
circumstellar dust \citep{1988ApJ...329L..93V, 1991ApJ...366..181M}.

\cite*{barniske2008} found a nebula around the  WN9h star
WR102ka in the Galactic Centre region.  They obtained a high-resolution
IR spectrum of the nebula and analysed the emission from warm dust and
molecules.   To our knowledge, this
was the first unambiguous detection of a dusty nebula around a
WN star \citep[cf.][]{2011ApJ...741....4F}.

Searches for circumstellar nebulae around evolved massive stars
with the old generation of IR telescopes (onboard the {\it
Infrared Astronomical Satellite}, the {\it Infrared Space
Observatory} and the {\it Midcourse Space Experiment satellite}),
led to the detection of a number of objects 
\citep[e.g.][]{1991ApJ...366..181M,1998Ap&SS.255..195T,2000A&A...356..501V,2002ApJ...572..288E,2003A&A...412..185C},
 some of which were already known from optical
and/or radio observations. Significant progress in detection of
new circumstellar nebulae was achieved with the release of the
MIPSGAL and other {\it Spitzer Space Telescope} Legacy Programmes.
Using {\it Spitzer} data, several groups independently discovered
many dozens of circumstellar shells 
\citep{2010MNRAS.405.1047G,2010AJ....139.1542M,Wachter}, while
follow-up spectroscopy of the central stars of these shells led to the
discovery of dozens of WR, luminous blue variable, and other 
massive stars 
\citep{2009MNRAS.400..524G,2010MNRAS.405.1047G,2010MNRAS.405..520G,2010MNRAS.403..760G,2011arXiv1110.0126G,Wachter,2011BSRSL..80..291W,Mauerhan,stringfellow20112012,2012IAUS..282..267S}.

Two WR stars revealed with {\it Spitzer} via detection of their
circumstellar shells are the subject of this paper. The WR nature
of these stars (formerly known as 2MASS J18420630-0348224 and
2MASS J18420827-351039) were identified by \cite{Mauerhan}
by means of IR spectroscopy. In the systematics of the Galactic WR
star catalogue \citep{hucht2001,hucht2006}\footnote{See updated version of this catalogue on http://pacrowther.staff.shef.ac.uk/WRcat}, we call these
stars WR120bb and WR120bc in the rest of the paper. \cite{Mauerhan} also identified two more evolved massive stars (of
spectral types of WC8 and O7$-$8\,III$-$I), which along with
WR120bb and WR120bc form a previously unknown star cluster, provided they lie at similar distances.

Analyses of WR stars are preferably based on their optical and
ultraviolet spectra, which are not accessible for WR120bb and
WR120bc because of high interstellar extinction. However, attempts to constrain the basic parameters of WR stars on basis of their $K$-band spectra alone have been made
\citep[e.g.][]{barniske2008,liermann2010}. In the present paper
the addition of the $J$ and $H$-bands are included along with the K-band analysis,
which turn out to be very valuable.

The rest of the paper is organised as follows. In Section\,2 we
introduce WR120bb and WR120bc and their circumstellar nebulae in
more detail. Section\,3 describes our spectroscopic observations.
After a short characterisation of the Potsdam Wolf-Rayet (PoWR)
model atmospheres, the spectra are quantitatively analysed in
Section\,4. The evolutionary status of the stars and the origin of
their nebulae are discussed in Section\,5. We summarise our findings in
Section\,6.

\section{The circumstellar shells MN85 and MN86 and their central stars}
\label{sec:neb}

\begin{figure*}
\begin{center}
\includegraphics[width=\textwidth,angle=0,clip=]{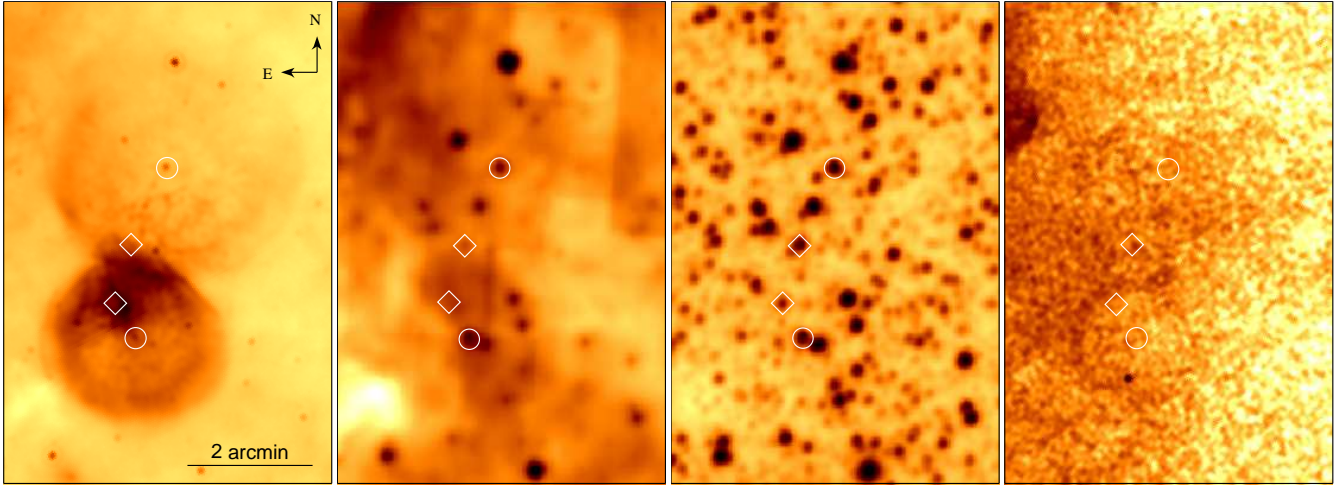}
\end{center}
\caption{From left to right: images at 24\,$\mu$m (MIPS), 12\,$\mu$m ({\it
WISE}), 3.4\,$\mu$m ({\it WISE}), and 20\,cm (VLA) of the field
containing the two interacting nebulae MN85 and MN86. The positions of
the central (WN9h) stars of the nebulae are marked by circles, while two
other  massive stars (WC8 and O7$-$8\,III$-$I)  are indicated by
diamonds (see text for details).} \label{fig:neb}
\end{figure*}

Fig.\,\ref{fig:neb} (left panel) shows the 24\,$\mu$m image of two
adjacent circular shells and their central stars (indicated by circles).
The shells were discovered in the course of our search for evolved
massive stars via detection of their circumstellar nebulae in the
archival data of the {\it Spitzer} Legacy Programmes \citep[for
motivation and the results of this search see][]{2010MNRAS.405.1047G}. The
image was obtained with the Multiband Imaging Photometer for {\it
Spitzer} \citep[MIPS;][]{rieke2004} within the framework of the 24 and 70
Micron Survey of the Inner Galactic Disc with MIPS 
\citep[MIPSGAL;][]{carey2009}. We have called the shells MN85 and
MN86\footnote{In the SIMBAD data base the shells are named [GKF2010]
MN85 and [GKF2010] MN86.} \citep{2010MNRAS.405.1047G}. One of the shells
(MN86) can also be seen at 70\,$\mu$m  \citep[see fig.\,1
in][]{Mauerhan}. Both shells were also covered by the Galactic
Legacy Infrared Mid-Plane Survey Extraordinaire 
\citep[GLIMPSE;][]{benjamin2003} carried out with the Infrared Array
Camera  \citep[IRAC;][]{fazio2004} and the Mid-Infrared All Sky Survey
carried out with the {\it Wide-field Infrared Survey Explorer} 
\citep[{\it WISE};][]{wright2010}. The first survey provides images at
3.6, 4.5, 5.8 and 8\,$\mu$m, and the second one at 3.4, 4.6, 12 and 22\,$\mu$m. The shells are invisible in the IRAC wavebands, but both can be
seen in the {\it WISE} 22\,$\mu$m image (not shown here). There is also
a hint of 12\,$\mu$m emission probably associated with the nebulae
(second left panel of Fig.\,\ref{fig:neb}). In Fig.\,\ref{fig:neb}
(second panel from the right) we also show the {\it WISE} 3.4\,$\mu$m
image of the same field, where the central stars of MN85 and MN86
are indicated by circles and the WC8 and O7-8 III-I stars (see Section\,1)
are marked by diamonds.

MN85 and MN86 were also covered by the Multi-Array Galactic Plane
Imaging Survey \citep[MAGPIS; ][]{helfand2006} carried out with the
Very Large Array (VLA). Fig.\,\ref{fig:neb} (right panel) shows that
the shell MN86 has a clear radio counterpart at 20\,cm.

The MIPS 24\,$\mu$m image shows that both shells are brightest where the two shells apparently overlap, which might be caused by the
interaction between the radially expanding shells \citep[cf.][]{2010MNRAS.405.1047G,Wachter}.
Indeed, \citet{Mauerhan} showed that the overlapping region of the shells is
significantly brighter than the sum of the surface brightnesses of
the neighbouring parts of the shells, which indicates that it
is intrinsically bright emission. The enhanced brightness of this region can
also be seen in the 70\,$\mu$m image of the shells in fig.\,1 in \citet{Mauerhan}.
If the shells are indeed interacting with
each other, then their almost circular shape implies that they came
in contact only recently. The existing data, however, do not allow
us to unambiguously assert that the physical contact between the
shells is real.

The angular radius of the larger shell (MN85) is $\simeq 1.9$\,arcmin,
while that of MN86 is $\simeq 1.5$\,arcmin. In Section\,\ref{sec:dist}
we argue that the distance to the objects is
$\sim 6$\,kpc, so that the linear radius of MN85 and MN86 is $\sim 3$\,pc. Circumstellar shells of this size are typical of late WN-type
(WNL) stars and luminous blue variables. The possible origin of MN85
and MN86 are discussed in Section\,\ref{sec:prog}.

The central location of point sources indicated in Fig.\,\ref{fig:neb}
by circles suggests that they might be associated with the nebulae. This
possibility is supported by the position of these stars in the
colour-colour diagrams by \cite{hadfield2007}, where they fall in the
region populated by WR stars. The details of the central stars are
summarised in Table\,\ref{tab:det}. The coordinates and the $J$, $H$,
$K_{\rm s}$ magnitudes are taken from the 2MASS (Two Micron All Sky
Survey) All-Sky Catalog of Point Sources \citep{cutri2003}. The $I$
magnitudes are from the DENIS (Deep Near Infrared Survey of the Southern
Sky) data base \citep{denis2005}, the IRAC magnitudes are from the
GLIMPSE Source Catalogue (I + II + 3D), and the {\it WISE} 12 and 22
$\mu$m magnitudes are from the {\it WISE} All-Sky Data Release \citep{cutri2012}. All these catalogues and data bases are
accessible through the Gator
engine\footnote{http://irsa.ipac.caltech.edu/applications/Gator/}.

\begin{table}
  \caption{Details of the central stars associated with the nebulae 
MN85 and MN86.}
  \label{tab:det}
  \begin{center}
 \begin{tabular}{lrr}
  \hline
    & WR\,120bb (MN85) & WR\,120bc (MN86) \\

  \hline
  RA(J2000) & $18^{\rm h} 42^{\rm m} 06\fs31$ & $18^{\rm h} 42^{\rm m} 08\fs27$ \\
  Dec(J2000) & $-03\degr 48\arcmin 22\farcs5$ & $-03\degr 51\arcmin 02\farcs9$ \\
  $l$ & $28\fdg4811$ & $28\fdg4452$ \\
  $b$ & $0\fdg3371$ & $0\fdg3094$ \\
  $I$ (mag) & 17.61$\pm$0.14 & 16.96$\pm$0.10 \\
  $J$ (mag) & 11.95$\pm$0.03 & 11.85$\pm$0.03 \\
  $H$ (mag) & 10.22$\pm$0.02 & 10.26$\pm$0.03 \\
  $K_{\rm s}$ (mag) & 9.16$\pm$0.02 & 9.27$\pm$0.03 \\
  $[3.6]$ (mag) & 8.28$\pm$0.04 & 8.43$\pm$0.04 \\
  $[4.5]$ (mag) & 7.63$\pm$0.04 & 7.88$\pm$0.05 \\
  $[5.8]$ (mag) & 7.43$\pm$0.03 & 7.67$\pm$0.03 \\
  $[8.0]$ (mag) & 7.04$\pm$0.03 & 7.20$\pm$0.03 \\
  $[12.0]$ (mag) & 6.92$\pm$0.04 & 6.96$\pm$0.04  \\
  $[22.0]$ (mag) & 5.23$\pm$0.05 & 3.85$\pm$0.06 \\
    \hline
 \end{tabular}
\end{center}
\end{table}

\section{Spectroscopic observations and data reduction}
\label{sec:obs}

Observations were carried out with the Apache Point Observatory (APO)
3.5m telescope  using  the medium-resolution infrared spectrograph
TripleSpec \citep{wilson2004},  which provides coverage from
~0.95 to 2.46\,$\mathrm{\mu m}$ in five orders with a resolving power of
about 3200 using a $1\farcs1 \times 43\arcsec$ slit.  The exposures 
were obtained in a nodded ABBA sequence offset along  the slit by
20\,$\mathrm{arcsec}$ to facilitate sky subtraction and to allow for the
correction of detector artifacts. The slit was aligned along the
east-west sky projection. WR\,120bb was  observed on 2010 May 19 UT at
an airmass of $\sim$1.55 with a seeing of $\sim$1\,$\mathrm{arcsec}$,
and WR\,120bc on 2010 May 30 UT at an airmass of 1.7  with
$\sim$1.2\,$\mathrm{arcsec}$ seeing.  Peak counts  in the strongest
emission lines for both stars remained below $\sim$14,000 counts, 
corresponding to a detector non-linearity of $<$\,0.1 per cent.

Flat fielding was performed using dome flat lamps, and wavelength
calibration utilized  the OH sky lines obtained during the exposures.
Sky subtraction, flat fielding,  wavelength calibration, and extraction
of the spectra were achieved using  a modified  version of the software
package {\tt xspextool} \citep*[originally developed for use with SpeX 
on the NASA Infrared Telescope Facility (IRTF); ][]{cushing2004}
redesigned for use with the APO TripleSpec specific instrument
characteristics. Final wavelength calibration in each order resulted in
an RMS error of $<\,$1.2 $\,\rm{\AA}$ for WR\,120bb, and $<$\,1.5\,$\rm{\AA}$
for  WR\,120bc. The A0\,V star HD\,171149 was observed immediately before
or after observing  each star for use in performing the telluric and
flux corrections. The difference in  airmass between the telluric and
programme stars were $<$\,0.06. Telluric corrections and approximate flux
calibrations were performed with the IDL routine {\tt xtellcor}  that
implements the method developed and described by \cite*{cushing2003}. 

The resulting combined spectrum of WR\,120bb has a typical
signal-to-noise ratio (S/N) in the continuum of $\approx 100$,
60, and 4$-$18 in the $K$, $H$, and $J$-bands, respectively.
Likewise, the values for the combined spectrum for WR\,120bc are 250,
200, and 6$-$75. The stars are heavily reddened, resulting in the
much lower S/N at shorter wavelengths.

\section{WR\,120\MakeLowercase{bb} and WR\,120\MakeLowercase{bc}: two
almost identical WN9\MakeLowercase{h} stars}

\subsection{Spectral type}
\label{sec:type}

\begin{figure*}
\begin{center}
\includegraphics[width=\textwidth,clip=]{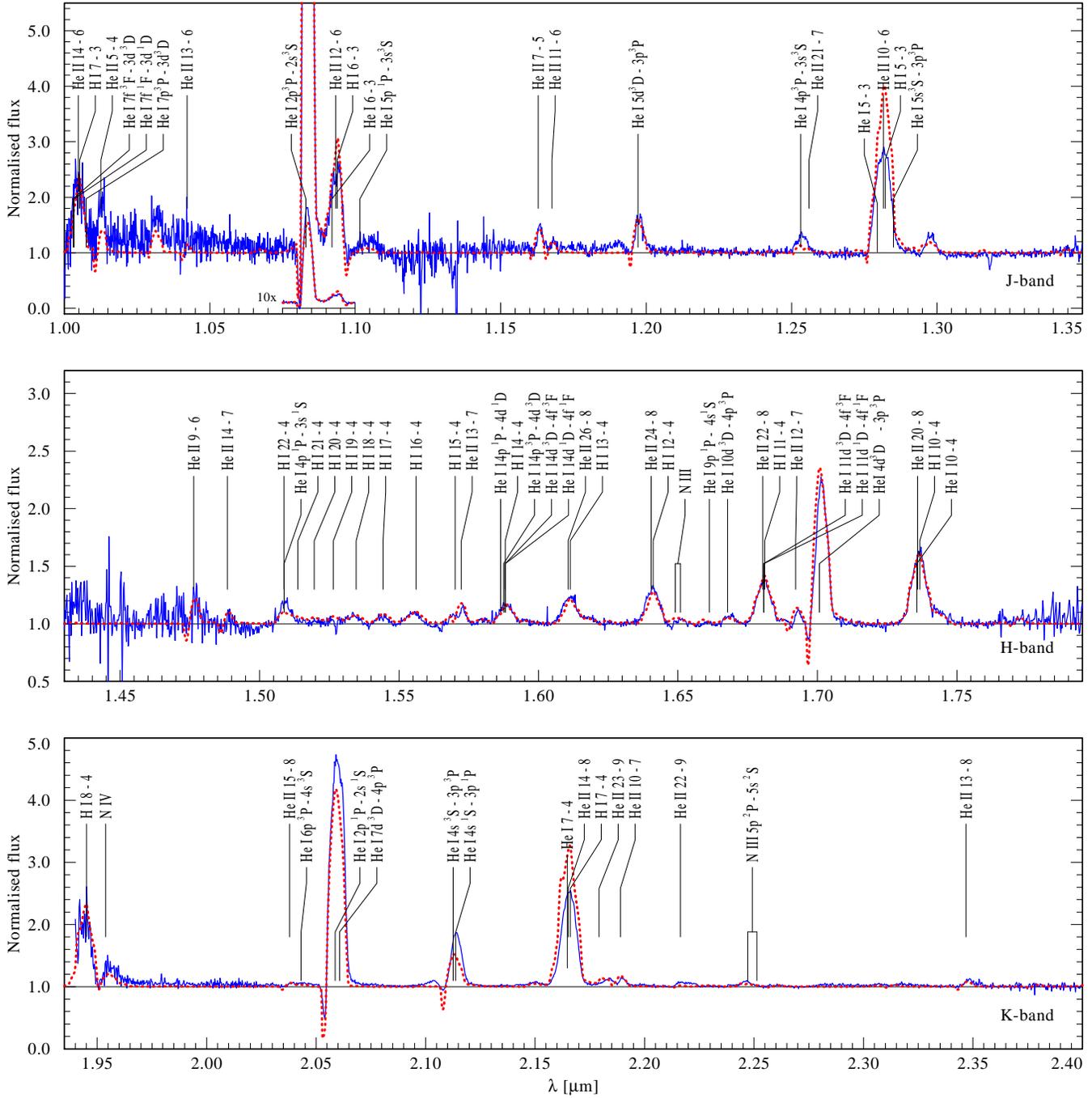}
\end{center}
\caption{Observed $J$, $H$, and $K$-band spectra (blue-solid line) of WR\,120bb,
compared with the best-fitting model (red-dotted line) with the
parameters as given in Table\,\ref{tab:model}. For normalisation the
observed spectra were divided by the reddened model
continuum, and then slightly aligned  ``by eye'' for an exact match of
the continuum slope.} 
\label{fig:linefita}
\end{figure*}

\begin{figure*}
\begin{center}
\includegraphics[width=\textwidth,clip=]{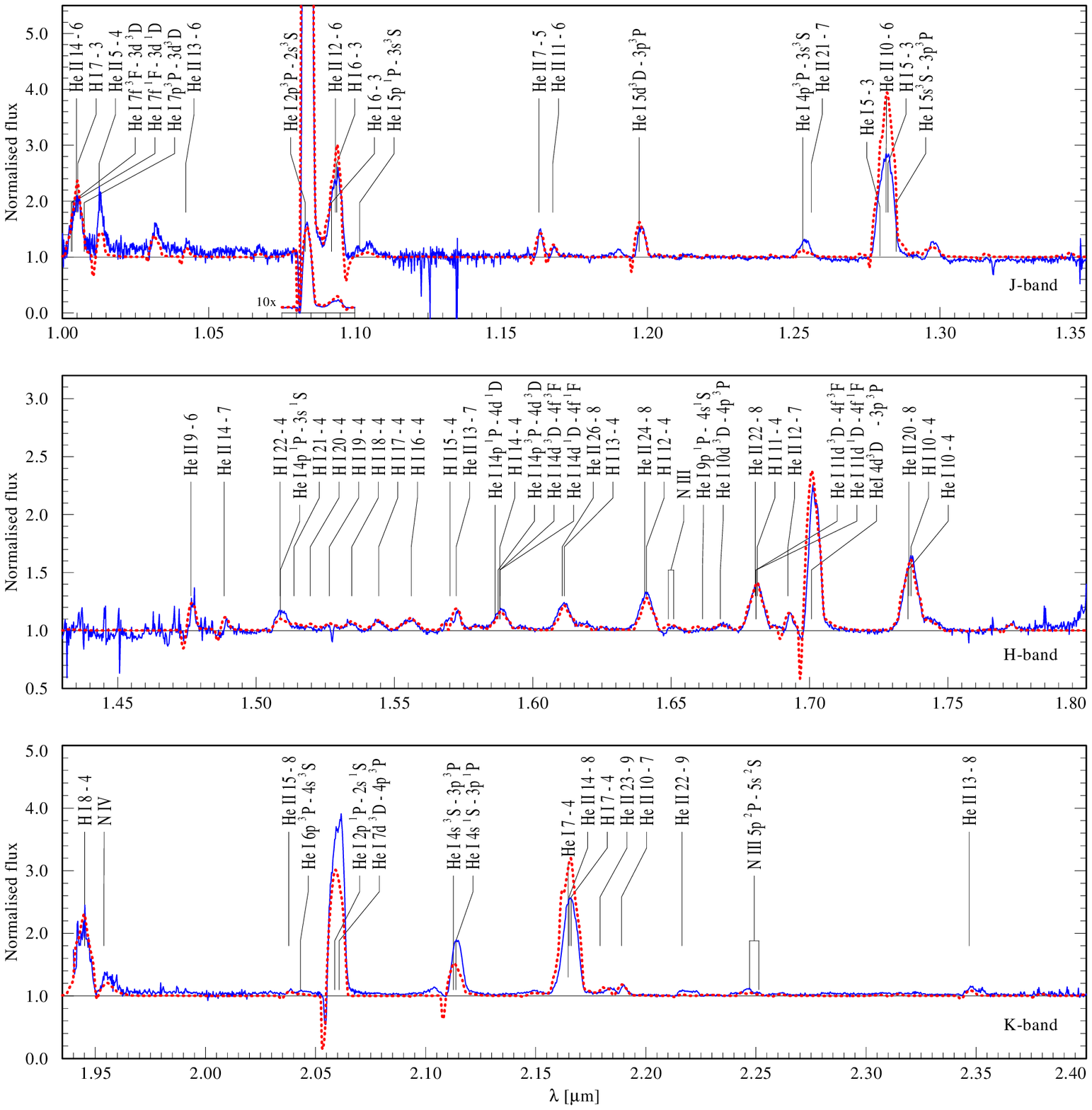}
\end{center}
\caption{Observed $J$, $H$, and $K$-band spectra (blue-solid line) of WR\,120bc,
compared with the best-fitting model (red-dotted line) as in
Fig.\,\ref{fig:linefita}}. 
\label{fig:linefitb}
\end{figure*}

In Figs.\,\ref{fig:linefita} and \ref{fig:linefitb} we present the $J$,
$H$, and $K$-band spectra of WR\,120bb and WR\,120bc. The spectra are
very similar to each other and show numerous emission lines of H\,{\sc
i}, He\,{\sc i} and He\,{\sc ii}. Hydrogen lines are generally blended
with the corresponding He\,{\sc ii} lines.  We also detect emissions of
N\,{\sc iii} and N\,{\sc iv}. The absence of carbon and oxygen lines
indicate that both stars belong to the nitrogen (WN) sequence of the WR
stars. To determine the subtypes of WR\,120bb and WR\,120bc more
precisely, we use the equivalent width (EW) ratios: EW(He\,{\sc ii}
1.012)/EW(He\,{\sc i} 1.083) and EW(He\,{\sc ii} 2.189)/EW(He\,{\sc ii}
+ Br$\gamma$) \citep[see][and Table\,5 therein]{crowther2006}. Following
\cite{crowther2006}, WN stars belong to the subtype WN8 if the
equivalent width ratio EW(He\,{\sc ii} 1.012)/EW(He\,{\sc i} 1.083) is
between 0.07 and 0.2 and if EW(He\,{\sc ii} 2.189)/EW(He\,{\sc ii} +
Br$\gamma$) is between 0.1 and 0.4. If the equivalent width ratios are
below these values, the stars belong to the subtype WN9.  If we compare
these ratios with the values that we found for WR\,120bb and WR\,120bc
(see Table \ref{tab:ew}), we can assign the two stars to the WN9
subtype. Following the three-dimensional classification for WN stars by
\cite*{smith1996}, we add a suffix `h' to WN9 to indicate the presence of
hydrogen emission lines, so that WR\,120bb and WR\,120bc are WN9h stars.
This result only slightly differs from that of \cite{Mauerhan}, who
assigned both stars to the WN8-9h subtype.

\begin{table}
\caption{Equivalent widths (EWs; in \AA) of emission lines and their ratios used
for spectral classification of WR\,120bb and WR\,120bc.} \label{tab:ew}
\begin{center}
\begin{minipage}{\textwidth}
\begin{tabular}{ccc}
\hline 
EW & WR\,120bb & WR\,120bc\\
\hline
He\,{\sc ii} 1.012 & 28.3$\pm$7.4 & 31.3$\pm$2.8\\
He\,{\sc i} 1.083 & 555.5$\pm$10.4 & 478.1$\pm$9.3\\
He\,{\sc ii} + Br$\gamma$ & 122.2$\pm$1.0 & 122.0$\pm$1.1\\
He\,{\sc ii} 2.189  & 3.4$\pm$0.3 & 5.0$\pm$0.2 \\
\hline 
EW(He\,{\sc ii} 1.012)/EW(He\,{\sc i} 1.083) & 0.051 & 0.065\\
EW(He\,{\sc ii} 2.189)/EW(He\,{\sc ii} + Br$\gamma$) & 0.028 & 0.041\\
\hline
\end{tabular}
\end{minipage}
\end{center}
\end{table}

\subsection{Spectral analysis and stellar parameters}
\label{sec:PoWR}

To analyse the stellar spectra and to derive the fundamental
parameters of WR\,120bb and WR\,120bc, the PoWR models for expanding stellar atmospheres are used.  These models
account for complex model atoms including iron-line blanketing in
non-LTE  \citep[for a detailed description see][]{graefener2002,hamann2003,hamann2004}. For the
ana\-lysis of the near-IR spectra, the model atom of H\,{\sc i}
had to be extended compared to the model grid for WNL-stars
\citep{hamann2004} in order to reproduce the Brackett-lines in the
$H$-band. Other extended model atoms are He\,{\sc i}, He\,{\sc ii},
N\,{\sc iii} and N\,{\sc iv}.
 The new atomic data are taken from the NIST Atomic Spectra Database \citep{nist} and from the Atomic Line List \citep{all}.

The normalised emission line spectra of WR stars depend on   two
principal parameters: the stellar temperature, $T_\ast$, and the
so-called transformed radius, $R_{\rm t}$.  The stellar temperature
$T_*$ denotes the effective temperature, which is related to the
stellar radius $R_{\ast}$ and the bolometric luminosity $L$ via the
Stefan-Boltzmann law \mbox{$L = 4 \pi R_\ast^2 \sigma T_\ast^4 $},
where $\sigma$ is the Stefan-Boltzmann constant and $R_\ast$ is by
definition the radius where the Rosseland optical depth reaches 20.
 Alternatively, one may quote the radius where the Rosseland mean optical depth of the stellar wind is $\tau_R = 2/3$. Since the winds of our two program stars are very dense, this ``pseudo-photosphere'' is located far out in the wind, at about $3.5\,R_{\star}$ (see Table \ref{tab:model}). Related to that radius, the effective temperature $T_{2/3}$ is much lower than $T_{\star}$ (see Table \ref{tab:model}).

The transformed radius $R_{\rm t}$ is related to the mass-loss rate $\dot{M}$ and defined by

\begin{eqnarray}
\label{eq:rtrans} 
R_{\rm t} = R_*  \left[\frac{v_\infty}{2500 \, \kms} \left/
  \frac{\sqrt{D} \dot M}
       {10^{-4} \, \msun \, {\rm yr^{-1}}}\right]^{2/3} \right. ~,
\end{eqnarray}

\noindent introducing the clumping contrast $D$ and the terminal velocity of the wind $v_\infty$.

In order to achieve a good fit of the electron scattering wings, we set
$D=10$. This is consistent with recent discussions of clumping
\citep*[see][]{hamann2008}. Note that the effect of $D$ is simply a
scaling of the empirically derived mass-loss rate, which is proportional to
$D^{-1/2}$. 

For the velocity field of the wind we adopt the usual
\mbox{$\beta$-law}, with the terminal velocity $v_{\infty}$ being a
free parameter. The exponent $\beta$ is set to unity throughout this
work.
 
The basic model parameters $T_\ast$ and $R_{\rm t}$ are derived from
fitting the lines of the model spectra to the lines in the normalised
observed spectra (see Figs.\,\ref{fig:linefita} and
\ref{fig:linefitb}). The normalisation is achieved by dividing the flux
calibrated observed spectra by the theoretical continuum.

For a systematic determination of those models which fit the
observations best, we calculated a small grid of models in the
$T_{\ast}-R_{\rm{t}}$ parameter plane.  
The terminal wind velocity is \mbox{$v_{\infty} =
750\,\kms$}, as inferred from the P-Cygni like line profiles of
the He\,{\sc i} lines at 1.70\,$\mathrm{\mu} \mathrm{m}$ (He\,{\sc i}
4d $^3$D - 3p $^3$P) and 2.06\,$\mathrm{\mu} \mathrm{m}$ (He\,{\sc i}
2p $^1$P - 2s $^1$S and He\,{\sc i} 7d $^3$D - 4p $^3$P). The
determination of the terminal wind velocity is considerably facilitated
by fitting both $H$- and $K$-band spectra.  Now we can use two lines with P-Cygni like profiles. As each of both lines gives a slightly different range for $v_{\infty}$, the real value can be constrained to the overlap of both intervals.
 
As the spectra of WR\,120bb and WR\,120bc look very similar
the best-fitting models have nearly identical parameters
(Table\,\ref{tab:model}).

The stellar temperature is adjusted such that the balance between the
lines from  He\,{\sc i} versus He\,{\sc ii} is reproduced. 
The He\,{\sc ii} 7-5 line at 1.16\,$\mathrm{\mu} \mathrm{m}$ in the $J$-band
and the He\,{\sc ii} 9-6 line at 1.48\,$\mathrm{\mu} \mathrm{m}$ in the
$H$-band are the most important indicators for the determination of the temperature. 
 
This is important to keep in mind, because there are other
parameter combinations with much lower temperatures which give well
fitting model spectra as well, but only models with high temperatures
of about $63\,\rm{kK}$ can reproduce the strengths of these He\,{\sc ii}
lines.

At the same time, $R_{\rm{t}}$ is adjusted, which influences the
strength of the emission lines in general.

The hydrogen to helium ratio is derived from fitting the line strengths
of the Brackett series in the $H$-band. As there are few strong hydrogen
lines in the $K$-band, the analysis of the $H$-band, which contains most of the
Brackett-series, was essential to determine the hydrogen abundance. For
the other elements, there were only weak and no unblended lines, so
that we could not determine their abundances. That is why for other
elements mass fractions which are typical for Galactic WN stars --
N:~1.5, C:~0.01, Fe:~0.14 per cent \citep{hamann2004} -- are adopted.

\begin{table}
  \caption{Stellar parameters for WR\,120bb and WR\,120bc.}
  \label{tab:model}
  \begin{center}
 \begin{tabular}{lrr}
  \hline
   Stellar parameters& WR\,120bb & WR\,120bc\\
  \hline
  Spectral type & WN9h & WN9h\\
  $T_{\ast}$ $\lbrack \mathrm{kK} \rbrack$ & 63 $ \pm$ 2 & 63 $ \pm$ 2\\
  $\log R_{t}$ $ \lbrack R_{\odot} \rbrack$ & 0.36 $ \pm$ 0.01 & 0.37 $ \pm$ 0.01\\
  $\log L$ $ \lbrack L_{\odot} \rbrack$ $^{\dagger}$  & 5.66 & 5.54\\
  $v_{\infty}$ $ \lbrack \kms \rbrack $& 750 $ \pm$  50 & 750 $\pm$ 50\\
  $D$ & 10 & 10\\
  $R_{\ast}$ $\lbrack R_{\odot} \rbrack$ $^{\dagger}$ & 5.7 & 4.9\\
  $\log \dot{M} $ $\lbrack M_{\odot} \mathrm{yr}^{-1} \rbrack $ $^{\dagger}$& $-$4.4 & $-$ 4.5\\
  $\log \Phi_i $ $\lbrack  \mathrm{s}^{-1} \rbrack $& 48.6 & 48.6\\
  $ R_{2/3}$ $\lbrack R_{\star} \rbrack$ & 3.6 & 3.4\\
  $ T_{2/3}$ $\lbrack \mathrm{kK} \rbrack$ & 33 $\pm$ 2 & 34 $\pm$ 2\\
  $E_{B-V}$ $ \lbrack \mathrm{mag} \rbrack $& 5.0 & 4.7\\
  $A_{K_{\mathrm{S}}}$ $ \lbrack \mathrm{mag} \rbrack $& 2.0 & 1.8\\
  $R_V$& 3.4  & 3.4\\
  $M_{K_{\mathrm{S}}} \lbrack \mathrm{mag} \rbrack $  & $-$ 6.7& $-$6.4 \\
  $d $ $\lbrack \mathrm{k} \mathrm{pc} \rbrack$ (adopted) &  5.8 &  5.8\\
  He [per cent by mass] & 73 $\pm$ 5 & 73 $\pm$ 5 \\
  H [per cent by mass] & 25 $\pm$ 5 & 25 $\pm$ 5 \\
  N [per cent by mass] & 1.5  & 1.5  \\
  C [per cent by mass] & 0.01   & 0.01 \\
  Fe [per cent by mass] & 0.14  & 0.14 \\
  \hline
 \end{tabular}
\begin{itemize}
 \item[$\dagger$] \footnotesize{ The error margins of $L$, $R_\ast$ and $\dot{M}$ due to the spectral analysis are small, but their values scale with the adopted distance $d$ which is less certain (see Sect. \ref{sec:dist}).}
\end{itemize}

\end{center}
\end{table}

\begin{figure}
\begin{center}
\includegraphics[width=\columnwidth,clip=]{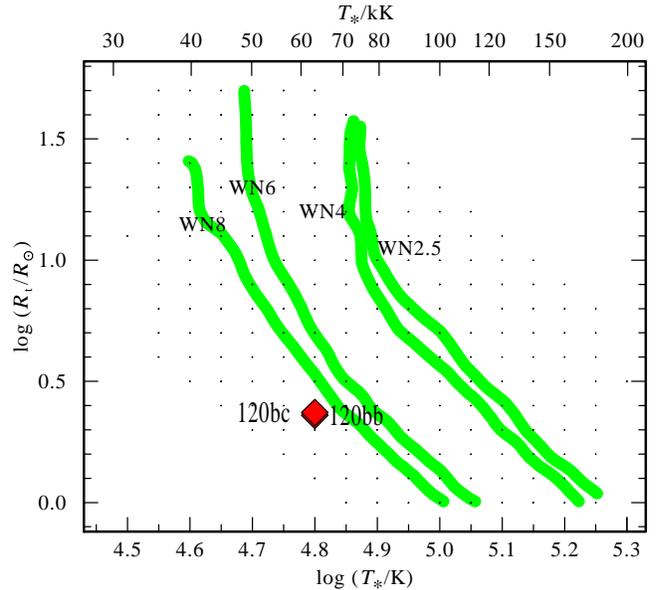}
\end{center}
\caption{The positions of WR\,120bb and WR\,120bc (overlapping diamonds) in the
$\log R_{\rm t} - \log T_\ast$ diagram. The thick lines indicate where the
classification criteria for selected subtypes (labels) match the line ratios
that are predicted from theoretical models. The underlying model grid was
calculated with the PoWR code \citep[see][for details]{hamann2004}.}

\label{fig:subtype}
\end{figure}

The parameters from the best-fitting model for WR\,120bb and WR\,120bc
are compiled in Table\,\ref{tab:model}. The stellar
temperature of 63\,kK is unusually high for subtypes as late as WN9.
As shown in Fig. \ref{fig:subtype}, the domains of the WN subtypes  in
the $T_{\ast}-R_{\rm{t}}$-plane have a slope. Usually, the WN stars
with late subtype and hydrogen are located at large $R_{\rm t}$, i.e.\
their winds are relatively thin \citep*[see][]{hamann2006}. Our programme
stars, however, show very  thick winds (small $R_{\rm t}$).  Their
location in the $T_{\ast}-R_{\rm{t}}$-diagram is consistent  with the
WN9 subtype derived above (Section\,\ref{sec:type}) which is based on
line {\em ratios}. However,  with the unusually high density in the
wind, higher stellar temperatures are required to reproduce the
observed amount of ionised helium. 

In addition, one should remember that $T_\ast$ refers by definition to the radius $R_\ast$ where the radial Rosseland optical depth reaches 20. This point of the atmosphere is of course not directly seen from outside. The observable photosphere is rather located at $R_{2/3}$ where the Rosseland optical depth has dropped to 2/3. The winds of our program stars are so thick that $R_{2/3} / R_\ast$ is as large as 3.6 and 3.4, respectively (see Table\,3). The effective temperature related to that radius is only $T_{2/3}$ = 33 or 34\,kK. Thus the location of $R_\ast$ depends on the model assumption that the $\beta$-law for the velocity continues in the optically thick part of the expanding atmosphere, albeit this cannot be observed, and has no influence on the spectral modelling.

\subsection{Luminosity and distance of WR\,120bb and WR\,120bc}
\label{sec:dist}

The absolute dimensions of the observed stars cannot be determined
by spectral analysis only. They are related to the distance of the
objects which is a priori unknown for WR\,120bb and WR\,120bc. Therefore,
additional assumptions about the distance or the absolute
magnitude have to be made in order to determine the luminosity,
the stellar radius and the mass-loss rate.

The sightline towards WR\,120bb and WR\,120bc ($l\approx 28\degr$) intersects
the Sagittarius Arm (located at $\sim 1.5$\,kpc from the Sun) and
then (at $\sim 3$\,kpc) enters into the Scutum-Centaurus (or
Scutum-Crux) Arm \citep[e.g.\ ][]{churchwell2009}. At a distance of
$\sim 6$\,kpc the sightline crosses a region of intense star
formation \citep{garzon1997}, which is located in the place
where the Scutum-Centaurus Arm meets the receding tips of the
Galactic traditional and long bars \citep{hammersley2000,cabrera2008,churchwell2009}.
This region
spans the Galactic latitude range from $\sim 24\degr$ to $32\degr$
and contains numerous very massive clusters of red supergiants
\citep{figer2006,davies2007,clark2009,alexander2009,negueruela2010,negueruela2011,gonzalez2012}
and the giant
H\,{\sc ii} region W43 with a deeply embedded ($A_V \sim 30$ mag)
cluster (named in the SIMBAD data base as [BDC99] W43 cluster),
which contains at least three evolved massive stars 
\citep*[the WN7+a/OB?
star WR 121a and two O-type giants or supergiants;][]{blum1999}
\footnote{The large ionising flux of the W43
star-forming region 
\citep*[$\sim 10^{51}$ Lyman continuum photons
$s^{-1}$;][]{smith1978} is comparable to that of
the very massive ($\sim 10^4 \, \msun$) star cluster NGC\,3603,
which suggests that the central cluster in W43 contains a large
number of yet undetected massive stars.}.
\begin{figure*}
\begin{center}
\includegraphics[width=18cm,angle=0,clip=]{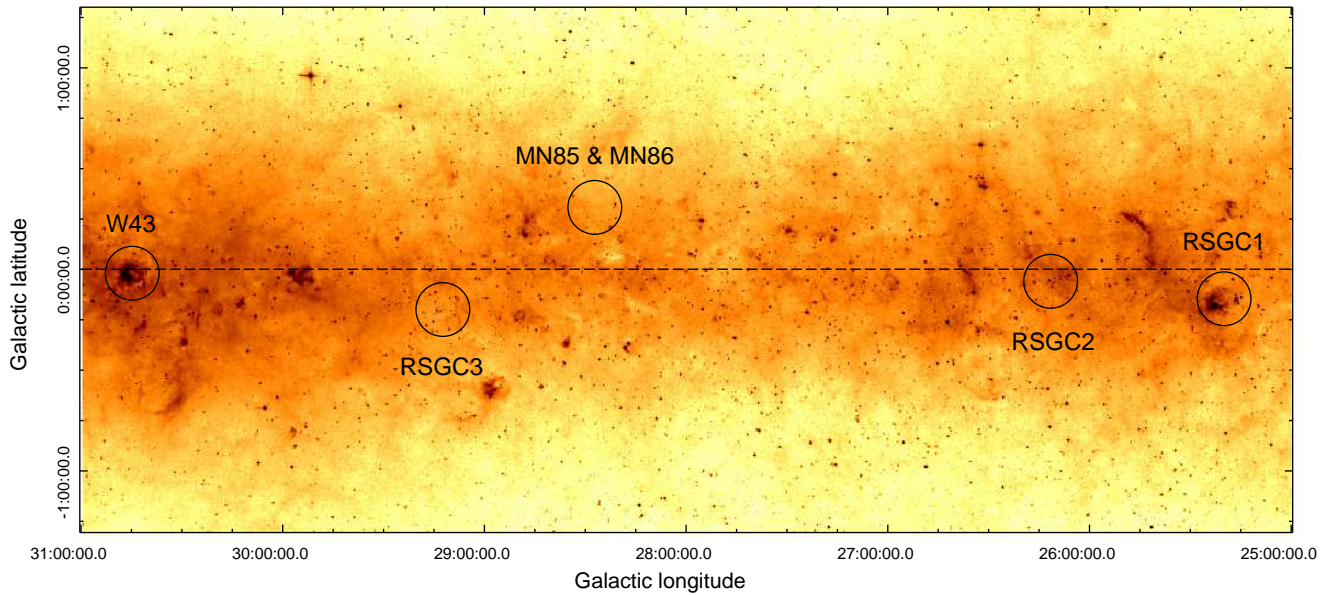}
\end{center}
\caption{{\it MSX} $8.3\,\mu$m image of the Galactic plane centred
at $l=28\degr, b=0\degr$. The positions of three red
supergiant clusters (RSGC1,2,3), the central cluster of the giant
H\,{\sc ii} region W43, and the cluster associated with WR\,120bb and
WR\,120bc are indicated by circles. The Galactic plane is shown by a
dashed line. See text for details.} \label{fig:RSGCs}
\end{figure*}
Fig.\,\ref{fig:RSGCs} shows the {\it Midcourse Space Experiment}
({\it MSX}) satellite \citep{price2001} image of a region of the
Galactic plane centred at $l=28\degr ,b=0\degr$, with the
positions of three red supergiant star clusters, RSGC1 \citep{figer2006}, RSGC2 \citep{davies2007}, and RSGC3 \citep{clark2009,alexander2009}, the central cluster of the giant H\,{\sc ii} region W43 \citep{blum1999}, 
and the cluster associated with WR120bb and WR120bc \citep{Mauerhan} indicated by circles.

To constrain the distance to WR\,120bb and WR\,120bc, we may adopt plausible values for their absolute magnitude in the $K$-band.  For eleven WN8-9h stars in the Arches cluster, located at the known distance of the Galactic center, \cite{martins2008} obtained a mean $K$ magnitude of $-6.5$\,mag with an individual scatter between $-6.8$ and $-5.8$\,mag\footnote{Note that these magnitudes were derived under the assumption of a uniform extinction across the Arches cluster of $A_K =2.8$ mag, while it actually ranges from $\approx 2$ to 4\,mag \citep*{espinoza2009}. Thus, the actual range of $M_K$ for WN8-9h stars remains unclear.} However, these WN stars in the Arches cluster might be not representative for the stars discussed here, since the former are very luminous, but have relatively low mass loss rates which reduces their IR excess. 

The calibration of $M_{K_{\rm s}}$ for Galactic WN stars by \citet{crowther2006} suggests for WN7-9 stars a much smaller value of $-5.9$\,mag. 
This value is the mean for only three stars, whose parent clusters 
and associations are believed to be known. It is, however, strongly 
biased by the WN8(h)+cc? star WR66, which is either located at a larger 
distance than implied from its possible membership in the anonymous association in Circinus \citep{lundstrom1984},
or is a special case of a low-luminosity WN star originating from the 
binary channel. Without this star, the mean $M_{K_{\rm s}}$ for the 
remaining two (WN7h and WN9h) stars is $-6.44$ mag, which agrees well 
with our choice.

Adopting $M_K =-6.5$\,mag and using the $K_{\rm s}$-band extinction derived 
from the spectral energy distribution (SED) fitting (see below), we get 
distances of $d=5.4$\,kpc for WR\,120bb and $d=6.2$\,kpc for WR\,120bc, 
which imply that the stars are located in the Scutum-Centaurus Arm and 
that the radial separation between them is 0.8\,kpc. Such a large 
separation, however, is less likely because both stars are probably members of 
a recently discovered star cluster \citep{Mauerhan} and therefore 
are most likely spatially associated with each other. 
As $M_K$ of the  WN8-9h stars in the Arches cluster span quite a wide range,
we assumed for both stars a common intermediate distance of  $d=5.8$\,kpc.
The corresponding stellar radii, mass-loss rates, luminosities, and 
absolute magnitudes are included in Table\,\ref{tab:model}. For the sake 
of completeness we also give in Table\,\ref{tab:model} the hydrogen 
ionising luminosity, $\Phi _{\rm i}$. 

Table\,\ref{tab:model} shows that the stellar parameters we 
obtained are very similar for both stars. However, the reddening towards WR\,120bb and 
WR\,120bc is quite different. This difference can be caused by the 
non-uniform extinction across the cluster, e.g. due to inhomogeneities 
in the remainder of the parent molecular cloud. Alternatively, the 
enhanced reddening could be caused by a dense clump of the circumstellar 
shell projected along the line of sight towards WR\,120bb \citep[cf.][]{gvaramadze2001}.

\begin{figure*}
\begin{center}
\includegraphics[width=\textwidth,clip=]{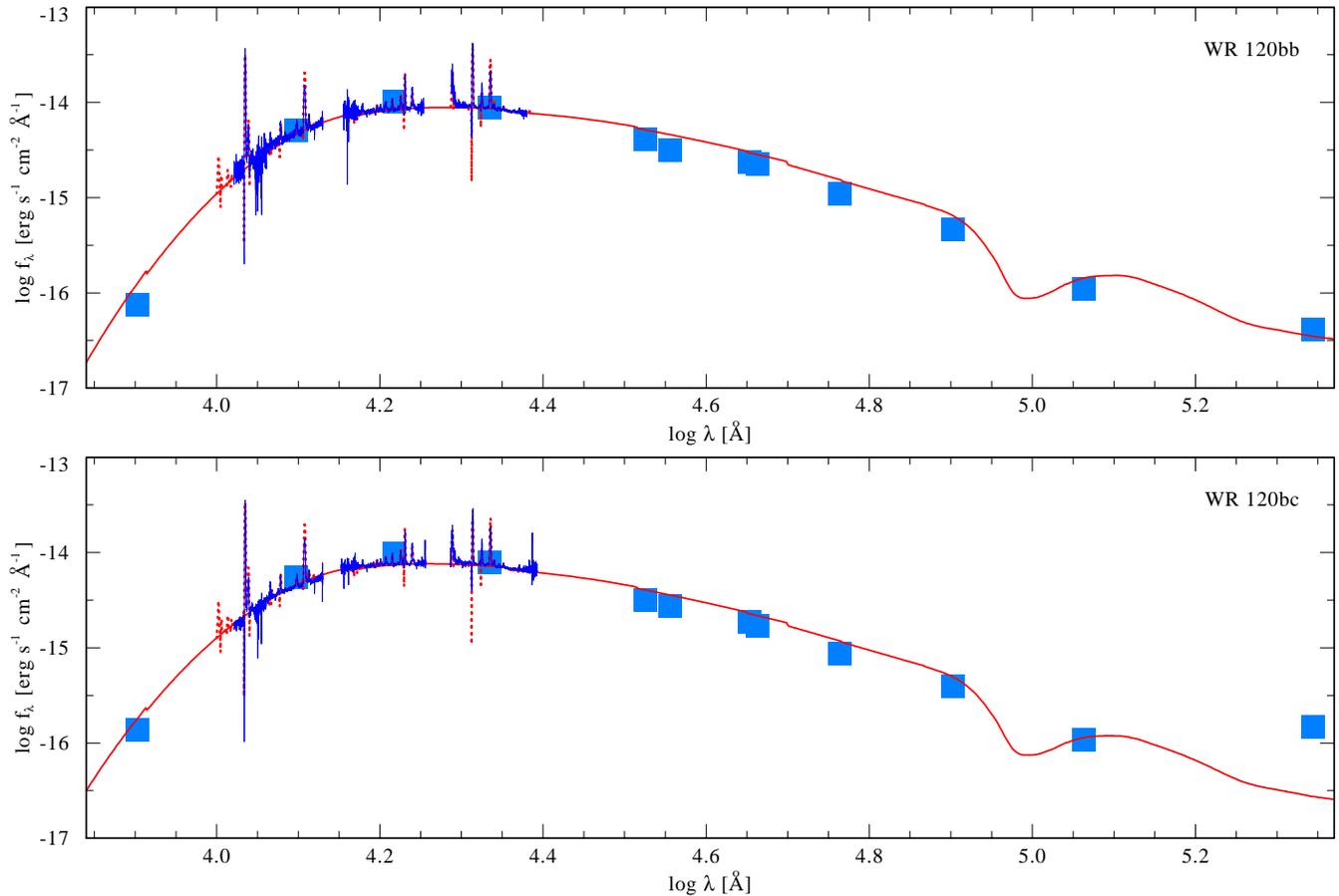}
\end{center}
\caption{Observed flux distribution of WR120bb and WR120bc (blue line) in absolute
units, including the calibrated spectrum and the photometric measurements compiled in
Table\,\ref{tab:det}, compared to the emergent flux of the model continuum (red/smooth line), in the
infrared also shown with spectral lines (red/dotted line). The model flux has been
reddened and scaled to the distance according to the parameters given in Table\,\ref{tab:model}.}
\label{fig:SEDa}
\end{figure*}

The observed SED was then fitted for each star (constructed from the
flux-calibrated spectra and the photometric observations from DENIS,
2MASS, IRAC and {\it WISE}) with the model SEDs (Fig.\,\ref{fig:SEDa})
by adjusting the parameters of the reddening curve.
For an appropriate treatment of the interstellar extinction, we adopted
the reddening law from \cite{Fitz}, which covers the 
wavelength range up to 50\,000\,$\rm \AA$ and allows adjustment of two parameters:
the colour excess $E_{B-V}$ and the total-to-selective absorption ratio
$R_V =A_V /E_{B-V}$, which is 3.1 for the diffuse interstellar medium
but can vary in general between 2.2 and 5.8 \citep{Fitz}. For wavelengths above 50\,000\,$\rm \AA$, the reddening law from  \cite{moneti} is used.  For the
distance of $d=5.8$\,kpc for both stars, good fits to their SEDs are
achieved with the parameters as given in Table\,\ref{tab:model}. 
The excess seen at 22\,$\mu$m obviously reflects the emission from dust, since the PSF  of the \textit{WISE} instrument (FWHM = 12$\arcsec$) covers a part of the circumstellar nebula.

The adopted distance to WR\,120bb and WR\,120bc of 5.8\,kpc implies that the parent cluster of these stars is located near the base of the Scutum-Centaurus Arm, i.e. in the region of active star formation caused by the interaction between the Galactic bar(s) and the Galactic disk \citep{garzon1997}. This inference is supported by a study of the distribution of interstellar extinction around $l\approx 28\degr$, which shows that $A_{K_{\rm s}}$ slowly increases from 1.2 to 1.6 between 5-6\,kpc, and then abruptly increases to $2-2.5$\,mag between 6 and 7\,kpc \citep[][and references therein]{negueruela2011}. Note that $A_{K_{\rm s}}$ towards WR\,120bb and WR\,120bc is comparable to that measured in the direction of the red supergiant clusters ($\approx 1.5 \pm 0.3$\,mag). Since it is believed that all these clusters are located at a common distance of $\sim 6$\,kpc \citep[e.g.\ ][]{clark2009,negueruela2011}, just in front of the large extinction wall, it is likely that the parent cluster of WR\,120bb and WR\,120bc is located at the same distance as well.

The actual distance to WR\,120bb and WR\,120bc, however, could differ from 5.8\,kpc. It would be somewhat smaller (for the given $M_K$) if the circumcluster 
medium significantly contributes to the reddening of these stars. In this 
connection we note that $A_{K_{\rm s}}$ towards the cluster embedded in W43 
(which is located at $d\sim 6$\,kpc) is $\sim 3$\,mag \citep{blum1999}.
Moreover, the absolute $K$-band magnitudes of WR\,120bb and WR\,120bc can differ from those given in Table\,\ref{tab:model}. Assuming that they are in the same range as those of the WN8-9h stars in the Arches cluster \citep[i.e. $-6.8 \ldots -5.8$\,mag;][]{martins2008}, one finds that the distance to WR\,120bb and WR\,120bc can range from $\sim 4.0$ to 7.0\,kpc, i.e. both stars are still located in the Scutum-Centaurus Arm. Correspondingly, the luminosities of WR\,120bb and WR\,120bc and their $R_{\ast}$ and $\dot{M}$ (given in Table\,\ref{tab:model}) can be scaled to different distances as $\propto d^2$, $\propto d$, and $\propto d^{3/2}$, respectively \citep*{schmutz1989}.

\section{Progenitors of WR\,120bb and WR\,120bc and the origin of their circumstellar nebulae}
\label{sec:disc}

\label{sec:prog}

The luminosity of WR\,120bb and WR\,120bc of $\log (L/L_{\odot}) \approx 5.5-5.7$ (derived for our preferred distance of 5.8\,kpc) implies that the initial (zero-age main-sequence) masses of these stars 
were $\la 40 \, M_{\odot}$ \citep[e.g.\ ][]{ekstrom2012}. A massive star of this mass
evolves through the sequence: MS (main sequence) $\rightarrow$ RSG
(red supergiant) $\rightarrow$ WN, and ends its life in a type II
supernova. During the MS phase the star creates an extended (tens
of pc) bubble filled with hot, tenuous gas \citep[e.g.][]{weaver1977}.
Then, during the RSG phase it loses a considerable fraction
of its initial mass in the form of slow, dense wind, which
occupies a compact (a few pc) region of enhanced density within
the MS bubble. This region is surrounded by a dense shell, which
is created by the high-pressure interior of
the MS bubble acting on the expanding RSG wind, so that the radius of the
shell, $r_{\rm RSG}$, is determined by the balance between the ram
pressure of the RSG wind and the thermal pressure in the interior
of the MS bubble \citep[e.g.\ ][]{ercole1992}:

\begin{equation}
{\dot{M}_{\rm RSG} v_{\rm RSG} \over 4\pi r_{\rm RSG} ^2} \approx
{7 \over (3850\pi )^{2/5}} \, L_{\rm MS} ^{2/5} \, \rho _0 ^{3/5}
\, t^{-4/5} \, , \label{eqn:ram}
\end{equation}
where $\dot{M} _{\rm RSG}$ and $v_{\rm RSG}$ are, respectively,
the mass-loss rate and the wind velocity during the RSG phase,
$L_{\rm MS}$ is the mechanical luminosity of the stellar wind
during the MS phase, $\rho_0 =1.4m_{\rm H} n_0$, $m_{\rm H}$ is
the mass of a hydrogen atom, and $n_0$ is the number density of
the ambient interstellar medium. Since WR\,120bb and WR\,120bc are members
of the cluster containing (at least) two other evolved massive
stars (whose progenitors were nearly as massive as those of WR\,120bb
and WR\,120bc), it is likely that all these stars contribute to the
thermal pressure of the common bubble around the cluster.

 Adopting $\dot{M} _{\rm RSG} =3\cdot 10^{-5} \msun \, {\rm yr}^{-1}$ and $v_{\rm RSG} =15\,\kms$ (the figures typical of RSGs), and assuming $L_{\rm MS} =3\cdot 10^{36} \, {\rm erg} \, {\rm s}^{-1}$ and $n_0 = 1 \, {\rm cm}^{-2}$, one has from equation\,(\ref{eqn:ram}) that at the 
beginning of the WR phase (i.e. at $t\ga 4$ Myr)
\begin{eqnarray}
r_{\rm RSG} \approx 3 \, {\rm pc} \left({\dot{M}_{\rm RSG} \over
10^{-5} \msun \, {\rm yr}^{-1}}\right)^{1/2} \left({v_{\rm RSG}
\over 15 \, \kms}\right)^{1/2} \nonumber \\
\cdot \left({L_{\rm MS} \over 3\cdot 10^{36} \, {\rm erg} \,
{\rm s}^{-1}}\right)^{-1/5} \left({n_0 \over 1 \, {\rm
cm}^{-3}}\right)^{-3/10} \left({t \over 4 \, {\rm
Myr}}\right)^{2/5} \, , \nonumber \label{eqn:rad-rsg}
\end{eqnarray}
which is comparable to the linear radii of the shells MN85 and
MN86 ($r_{\rm sh} \sim 3$\,pc; see Section\,\ref{sec:neb}).

The subsequent fast WR wind sweeps up the RSG wind and creates a
new shell, which expands through the region occupied by the RSG
wind with a constant velocity \citep[e.g.][]{chevalier1983,ercole1992}

\begin{equation}
v_{\rm sh} \approx \left({\dot{M} _{\rm WR} v_{\rm WR} ^2 v_{\rm
RSG} \over 3\dot{M} _{\rm RSG}}\right)^{1/3} \, , \label{eqn:WR}
\end{equation}
where $\dot{M} _{\rm WR}$ and $v_{\rm WR}$ are, respectively, the
mass-loss rate and the wind velocity during the WR phase. With
$\dot{M} _{\rm WR}$ and $v_{\rm WR}$ from Table\,\ref{tab:model},
one has from equation\,(\ref{eqn:WR}) $v_{\rm sh} \approx 160 \,
\kms$, which in turn gives the dynamical age of the shells of
$t_{\rm dyn} \sim r_{\rm sh} /v_{\rm sh} \sim 2\cdot 10^4$\,yr.
Note that $v_{\rm sh}$ and $t_{\rm dyn}$ only weakly depend on the
distance (both $\propto d^{1/2}$). The small dynamical age of the
shells implies that WR\,120bb and WR\,120bc entered the WR
phase only recently, which is consistent with the observational fact that
WRcircumstellar nebulae are associated exclusively with WNL stars
\citep[][and references therein]{2009MNRAS.400..524G,2010MNRAS.405.1047G}, i.e.
with very young WR stars, whose winds still interact with the
dense circumstellar medium. On the other hand, the small size of
the region occupied by the RSG wind implies that the circumstellar
shells around WR stars are short-lived 
\citep*[several tens of thousands of years; e.g.][]{marle2005} objects.

\section{Summary}
\label{sec:sum}
We analysed new $J$, $H$, and $K$-band spectra of two neighbouring
optically obscured Wolf-Rayet stars, WR\,120bb and WR\,120bc,
which were revealed via detection of their  (apparently interacting)
circumstellar shells with the {\it Spitzer Space Telescope} and
follow-up spectroscopy of central stars of the shells. Our
analysis of the spectra was based on the use of the Potsdam Wolf-Rayet model
atmosphere code, whose atomic data base has been extended in order
to include all significant lines in the near-infrared bands.
It is shown that the use of the $J$- and $H$-band is of great additional value for the analysis compared to an analysis which is based on the $K$-band spectrum alone.
Despite the late spectral subtype of WR\,120bb and WR\,120bc, we
found relatively high stellar temperatures of 63\,kK for both stars.
The stellar wind composition is dominated by helium with 25 per
cent of hydrogen. The stellar spectra are significantly reddened, 
$A_{K_{\rm s}} \approx 2$ mag,
which is consistent with the location of the stars
in the Scutum-Centaurus arm. 
Adopting a common distance of 5.8\,kpc, WR\,120bb and WR\,120bc have luminosities of $\log(L/L_{\odot} )=5.66$ and 5.54 and mass-loss rates of 
$10^{-4.4}$ and $10^{-4.5} \, \msun \, {\rm yr}^{-1}$, respectively.  These values have to be considered as highly uncertain due to the uncertainty of the adopted distance.
The inferred luminosities imply that the immediate precursors of
WR\,120bb and WR\,120bc were red supergiant stars of initial mass of  $\la 40 \, \msun$, which in turn implies that the circumstellar shells around both stars originate because of interaction between the Wolf-Rayet wind and the dense material shed during the preceding red supergiant phase.


\section{Acknowledgements}

We are grateful to the referee for a careful reading of the 
manuscript and comments that allowed us to improve the content of the paper. 
Observations presented were obtained with the Apache Point Observatory 3.5-meter telescope, which is owned and operated by the Astrophysical Research Consortium. 
A.Y.K acknowledges the support from the National Research Foundation
(NRF) of South Africa.
This work has made use of the NASA/IPAC Infrared Science Archive,
which is operated by the Jet Propulsion Laboratory, California
Institute of Technology, under contract with the National
Aeronautics and Space Administration, the SIMBAD database and the
VizieR catalog access tool, both operated at CDS, Strasbourg,
France.

\bibliographystyle{mn2e.bst}
\bibliography{aamnem99,literatur}

\label{lastpage}

\end{document}